\documentclass[a4paper, 11pt]{article}
\usepackage[left=19mm, right=19mm, bottom=35mm, top=35mm]{geometry}
\pdfoutput=1
\pdftrue

\usepackage{amsmath}
\usepackage[pdftex]{graphicx}

\usepackage[T2A]{fontenc}
\usepackage[utf8]{inputenc}
\usepackage[english]{babel}
\usepackage{multicol}
\usepackage{colortbl} 

\usepackage{geometry}
\usepackage{amsfonts}
\usepackage{amsmath}
\usepackage{amssymb}
\usepackage{ntheorem}
\usepackage{latexsym}
\usepackage{units}
\usepackage{textcomp}
\usepackage{wrapfig}
\usepackage{booktabs}
\usepackage{hyperref}

\usepackage{xcolor}
\hypersetup{
    colorlinks,
    linkcolor={red!50!black},
    citecolor={blue!50!black},
    urlcolor={blue!80!black}
}

\everymath{\displaystyle}
\usepackage{color}
\everymath{\displaystyle}

\renewcommand{\phi}{\varphi}

\usepackage{url}

\usepackage{sectsty} 
\sectionfont{\fontsize{11}{14}\selectfont}     
\subsectionfont{\fontsize{10}{13}\selectfont}  

\begin{document}


\vspace{-20mm}
\begin{center}
\section*{\Large Observation of \boldmath{$K^{+} \to \pi^{+}\pi^{0}\pi^{0}\gamma$} ~decay}
\end{center}
\vspace{2mm}
\begin{center}
\begin{minipage}{1.0\linewidth}
  {\center \large \textsc{The OKA collaboration}\\}\vspace{-2mm}
\end{minipage}
\end{center}
\begin{center}
\begin{minipage}{1.0\linewidth}
 \center{
  \textsc
  A.~V.~Kulik{${}^{a}$},
  S.~N.~Filippov,
  E.~N.~Gushchin,
  A.~A.~Khudyakov,\\
  V.~I.~Kravtsov,
  Yu.~G.~Kudenko{${}^{b,c}$},
  A.~Yu.~Polyarush,
 }\vspace{-3mm}
 \center{\small 
   \textsc{(Institute for Nuclear Research -- Russian Academy of Sciences, 117312 Moscow, Russia),} 
 }\vspace{-1mm}
 \center{
  \textsc
  A.~V.~Artamonov,
  S.~V.~Donskov,
  A.~P.~Filin,
  A.~M.~Gorin,
  A.~V.~Inyakin,
  G.~V.~Khaustov,\\
  S.~A.~Kholodenko,
  V.~N.~Kolosov,
  V.~F.~Kurshetsov,
  V.~A.~Lishin,
  M.~V.~Medynsky,\\
  V.~F.~Obraztsov,
  A.~V.~Okhotnikov,
  V.~A.~Polyakov,
  V.~I.~Romanovsky,
  V.~I.~Rykalin,\\
  A.~S.~Sadovsky,
  V.~D.~Samoilenko,
  I.~S.~Tiurin,
  V.~A.~Uvarov,
  O.~P.~Yushchenko
 }\vspace{-3mm}
 \center{\small 
   \textsc{(NRC "Kurchatov Institute"${}^{}_{}{}^{}$-${}^{}_{}{}^{}$IHEP, 142281 Protvino, Russia),} 
 }\vspace{-1mm}
 \center{
  \textsc
  V.~N.~Bychkov, 
  G.~D.~Kekelidze,
  V.~M.~Lysan,
  B.~Zh.~Zalikhanov
 }\vspace{-3mm}
 \center{\small 
   \textsc{(Joint Institute of Nuclear Research, 141980 Dubna, Russia)}\\
 }\vspace{-1mm}
\end{minipage}
\end{center}

{
\footnotesize
\line(1,0){170}\\
\vspace{-1mm}${}$\hspace{0.8cm}${}^{a}$~e-mail: alexkulik@ihep.ru\\ 
\vspace{-1mm}${}$\hspace{0.8cm}${}^{b}$~Also at National Research Nuclear University (MEPhI), Moscow, Russia\\
\vspace{-1mm}${}$\hspace{0.8cm}${}^{c}$~Also at Institute of Physics and Technology, Moscow, Russia\\
}
\vspace{-4mm}
\begin{center}
\begin{minipage}{0.09\linewidth}
~
\end{minipage} 
\begin{center}
\begin{minipage}{0.83\linewidth}
{ 
  \rmfamily
  {\bf Abstract}
  The $K^{+} \to \pi^{+}\pi^{0}\pi^{0}\gamma$ decay is observed by the OKA collaboration. The branching ratio is measured to be
  $(4.1 \pm 0.9(stat) \pm 0.4(syst))\times 10^{-6}$. The branching ratio and $\gamma$ energy spectrum are consistent with ChPT prediction.
}\vspace{3mm}
{\\  {\bf Keywords} {Kaon decays~$\cdot$~experimental results}}
\end{minipage}
\end{center}
\begin{minipage}{0.09\linewidth}
~
\end{minipage}
\end{center}
\vspace{-2mm}



\section{Introduction}\label{SectInitro}

The analysis of the $K^{+} \to \pi^{+}\pi^{0}\pi^{0}\gamma$ decay within the framework of Chiral Perturbation Theory was performed in

\cite{bibBR},\cite{bibMatrEl} predicting branching ratio at $3.76 \times 10^{-6} (E_{\gamma}^*>10)$MeV.
So far the observation is claimed by the single experiment \cite{Bolotov} with statistics of 5 events and
$BR=(7.6_{-3.0}^{+6.0}) \times 10^{-6}$.
In this article we present the observation and measurement of that decay with considerably improved precision.

\section{The OKA setup}
The {\bf OKA} is a fixed target experiment dedicated to the study of kaon decays using the decay in flight technique. 
It is located at NRC ''Kurchatov Institute''-$^{}$IHEP in Protvino (Russia). 
A secondary kaon-enriched hadron beam is obtained by an RF separation with the Panofsky scheme.
The beam is optimized for the momentum of 17.7~GeV/c with kaon content of about 12.5\% and intensity up to $5\times 10^{5}$ kaons per
U-70 accelerator spill.
\begin{figure*}[!ht]
\begin{center}
\includegraphics[width=1.00\textwidth]{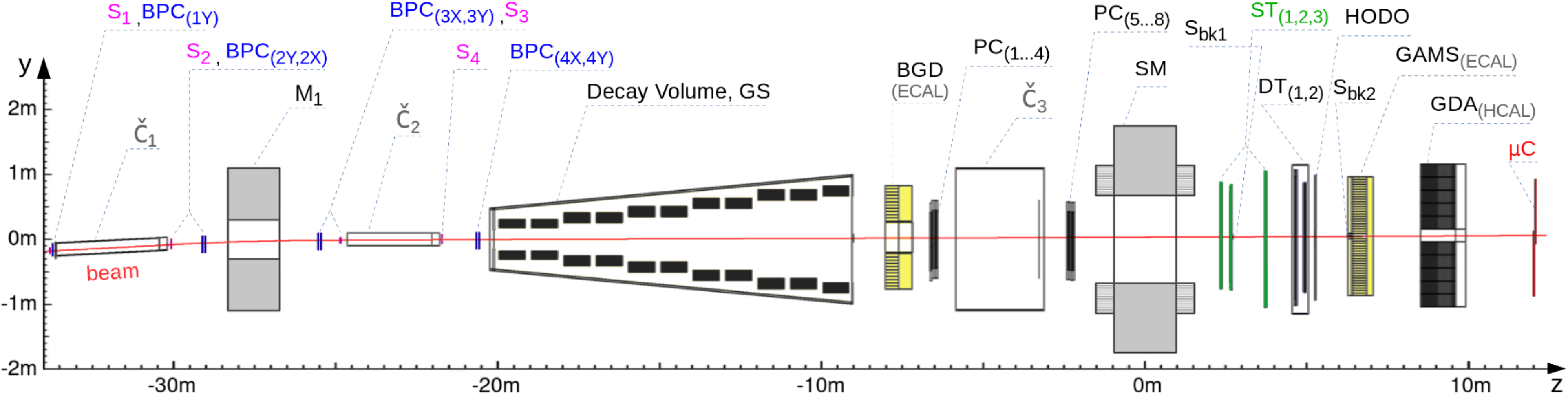}
\caption{\label{FigOkaSetup}Schematic elevation view of the OKA setup. See the text for details.}
\end{center}
\end{figure*}

The OKA setup (Fig.~\ref{FigOkaSetup}) makes use of two magnetic spectrometers along with an 11~m long {\small Decay Volume (DV)} filled with helium at atmospheric pressure 
and equipped with a guard system ({\small GS}) of lead-scintillator sandwiches mounted in 11 rings inside the {\small (DV)} for photon veto.
It is complemented by an electromagnetic calorimeter {\small BGD} \cite{BGDref1982} with a wide central opening.

The first magnetic spectrometer measures momentum of the beam particles with a resolution of $\sigma_{p}/p$ $\sim 0.8\%$. 
It consists of a vertically (y) deflecting magnet {\small M$_{1}$} surrounded by a set of 1~mm pitch (beam) proportional chambers {\small BPC${}_{(}$${}_{1Y,}$\hspace{-1pt} ${}_{2Y,}$\hspace{-1pt} ${}_{2X,...,}$\hspace{-1pt} ${}_{4Y}$${}_{)}$}.  It is supplemented by two threshold
Cherenkov counters {\small \v{C}$_{1}$,\v{C}$_{2}$} for kaon identification and by beam trigger scintillation counters 
{\small S$_{(1)}$}, {\small  S$_{(2)}$}, {\small S$_{(4)}$}.
To measure the charged tracks from decay products the second magnetic spectrometer is used (with resolution of $\sigma_{p}/p \sim$ 1.3 -- 2\% for momentum range of 2--14 GeV/$c$). 
It consists of a wide aperture 200$\times$140 cm$^2$ horizontally (x) deflecting magnet {\small SM} with $\int Bdl \sim 1$~Tm surrounded by tracking stations:
proportional chambers {\small PC$_{(1,...,8)}$}, straw tubes {\small ST$_{(1,2,3)}$} and drift tubes {\small DT$_{(1,2)}$}.
A matrix hodoscope {\small HODO} is used to improve time resolution 
and to link $x$--$y$ projections of a track. 

At the end of the setup there are: an electromagnetic calorimeter {\small GAMS$_{\tt(ECAL)}$} of 15X${}_{0}$ (consisting of $\sim$ 2300 \\ 3.8$\times$3.8$\times$45~cm$^3$ lead glass blocks) \cite{GAMSref1985},
a hadron calorimeter {\small GDA$_{\tt(HCAL)}$} of 5$\lambda$ (constructed from 120 20$\times$20~cm${}^{2}$ iron-scintillator sandwiches with WLS plates readout)
and a wall of  
$4\times1$m${}^{2}$ muon counters {\small {$\mu$}C} behind the hadron calorimeter. 

More details on the OKA setup can be found in \cite{OKA_status_2009, OKA_KmuHnu_EPJC}.

\section{The data and the analysis procedure}
Two sequential runs of data with beam momentum of 17.7\hspace{-0.5pt} GeV/c recorded by OKA collaboration in 2012 and in 2013  are analysed
in search for $K^{+} \to \pi^{+}\pi^{0}\pi^{0}\gamma$ decay.

The main trigger requires a coincidence of 4 beam scintillation counters, a combination
of two Cerenkov's ({\small \v{C}$_{1}$} sees pions, {\small \v{C}$_{2}$} pions and kaons) and, finally, anticoincidence
of two scintillation counters {\small S$_{bk1}$, S$_{bk2}$}, located on the beam axis after the magnet to suppress undecayed
beam particles:\\  ${\tt Tr_{Kdecay}=}$ ${\tt S_{1}{\cdot}S_{2}{\cdot}S_{3}{\cdot}S_{4}{\cdot}\overline{\check{C}}_{1}{\cdot}\check{C}_{2}{\cdot}\overline{S}_{bk} }$.
The trigger additionally requires an energy deposition in GAMS-2000 e.m. calorimeter higher than ~2.5 GeV 
to suppress the dominating $K^{+}\to\mu^{+}\nu$ decay: \\
${\tt Tr_{GAMS}=Tr_{Kdecay} \cdot (E_{GAMS}>2.5}$~GeV$)$.

The Monte Carlo (MC) statistics is generated with Geant-3.21 \cite{Geant321} program comprising a realistic description of the setup.
The MC events are passed through full OKA reconstruction procedures.

For the estimation of the background to the selected data set, a sample of the Monte Carlo events with six main decay channels of charged kaon
($\pi^{+}\pi^{0}$, $\pi^{+}\pi^{0}\pi^{0}$, $\pi^{+}\pi^{0}\gamma$, $\mu^{+}\nu\gamma$, $\pi^{0}\mu^{+}\nu$, $\pi^{0}e^{+}\nu$)
mixed accordingly to their branching fractions, with the total statistics
$\sim$ 10 times larger than that of the recorded data sample is used.
Every MC event has a weight $w \sim |M|^2$ where M is the matrix element of the decay.
The weights for the 3-body decays are calculated from the data, presented in PDG \cite{PDG}, the matrix element of the
$K^{+} \to \pi^{+}\pi^{0}\pi^{0}\gamma$ decay comes from \cite{bibMatrEl}. The processes when the beam kaon scatters or interacts
while passing the setup are also added.

\subsection{Event selection}
The total of $\sim$ $3.65\times 10^{9}$ events with kaon decays are logged
in of which $\sim$ $8\times 10^{8}$ events are reconstructed with a single charged particle in the final state.
The primary selection criteria are:

\begin{itemize}
  \item The angle between the beam and secondary tracks $\Theta>2mrad$ and the distance $<1cm$.
  \item The decay vertex is within the decay volume.
  \item Only one segment of the charged track downstream the analysing magnet.
  \item 5 $\gamma$ with energy $E > 0.5GeV$ detected.
  \item Out of all possible combinations of 4 $\gamma$s the one with minimal value of:\\
     $R^2 = (m_{ij} - m_{\pi^0})^2 + (m_{kl} - m_{\pi^0})^2,
    \quad ij\neq kl$ is identified as $\pi^0\pi^0$\\ and the remaining $5^{th}$ ``stray'' $\gamma $ considered bremsstrahlung.
  \item $5^{th} \gamma$ hits GAMS rather than BGD.
  \item At least one out of 4 $\gamma$, making $\pi^0\pi^0$, hits GAMS. This has some influence since GAMS is in trigger and BGD is not.
 \item $5^{th} \gamma$ energy in $K^+$ rest frame is $10<E_{\gamma}^*<50MeV$. 
\end{itemize}

\begin{figure}[!ht]
\begin{center}
\includegraphics[width=0.35\textwidth]{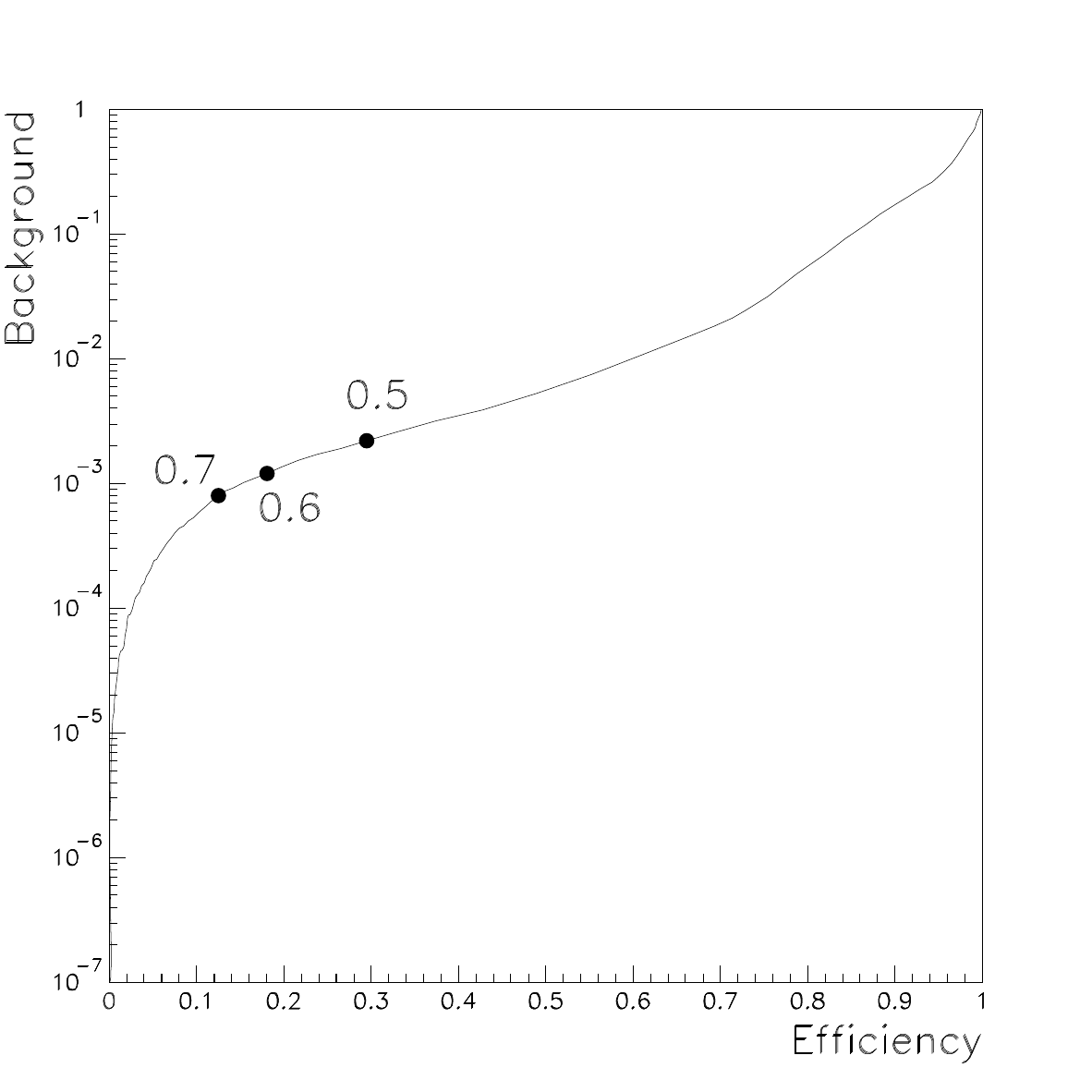}
\caption{\label{roc} Neural net performance, the thresholds used in this analysis shown with bullets.}
\end{center}
\end{figure}

About 230k events selected at this stage. Per MC simulation the only background process surviving this selection is non-radiative
decay $K^{+} \to \pi^{+}\pi^{0}\pi^{0}$ of similar topology less one $\gamma$. The extra ``ghost'' $\gamma$ easily emerges due to
the fluctuations of $\pi^+$ hadronic shower in GAMS e.m. cal. This background decay is $\approx 5000$ times more frequent ($1.76\%$) than the
radiative decay in question thus presenting a major challenge in this analysis. We employ a RBFN neural network\cite{RBFN} to suppress this
background. The neural network (NN) is trained on MC events of $K^{+} \to \pi^{+}\pi^{0}\pi^{0}$ and $K^{+} \to \pi^{+}\pi^{0}\pi^{0}\gamma$
decays, 100k of each. For the inputs to the NN we use the variables relevant to discrimination between two types of events:

\begin{itemize}

\item $\Delta E = E_{\pi^+} + \sum_{i=1}^{5} E_{\gamma_i} - E_{beam}$ - energy balance in the event. 

  \item $E_{\gamma}$ - the $5^{th} \gamma$ energy.
  
  \item $d_{\gamma}$ - distance from the $ 5^{th} \gamma$ to the track at GAMS plane.

  \item $\chi_{\gamma}^2$ - the $5^{th} \gamma$ $ \chi^2$ of shower shape fit.

  \item $\chi_{\pi^+\pi^0\pi^0}^2$ - $\chi^2$ of 3C-fit to $\pi^+\pi^0\pi^0$.

  \item $\chi_{\pi^+\pi^0\pi^0\gamma}^2$ - $\chi^2$ of 3C-fit to $\pi^+\pi^0\pi^0\gamma$.

  \item $M_{\pi^+\pi^0\pi^0}$ - mass of $\pi^+\pi^0\pi^0$ in 3C-fit 
\end{itemize}
 On its output the RBFN produces a real number. Moving the output threshold offers control over signal:background
ratio (fig.\ref{roc}).

Clear peak is seen in the mass spectra for 3 different thresholds on RBFN output (fig.\ref{bkg-sub}).

\subsection{Fitting mass spectra}

\begin{figure*}[!ht]
\begin{center}
\includegraphics[width=0.3\textwidth]{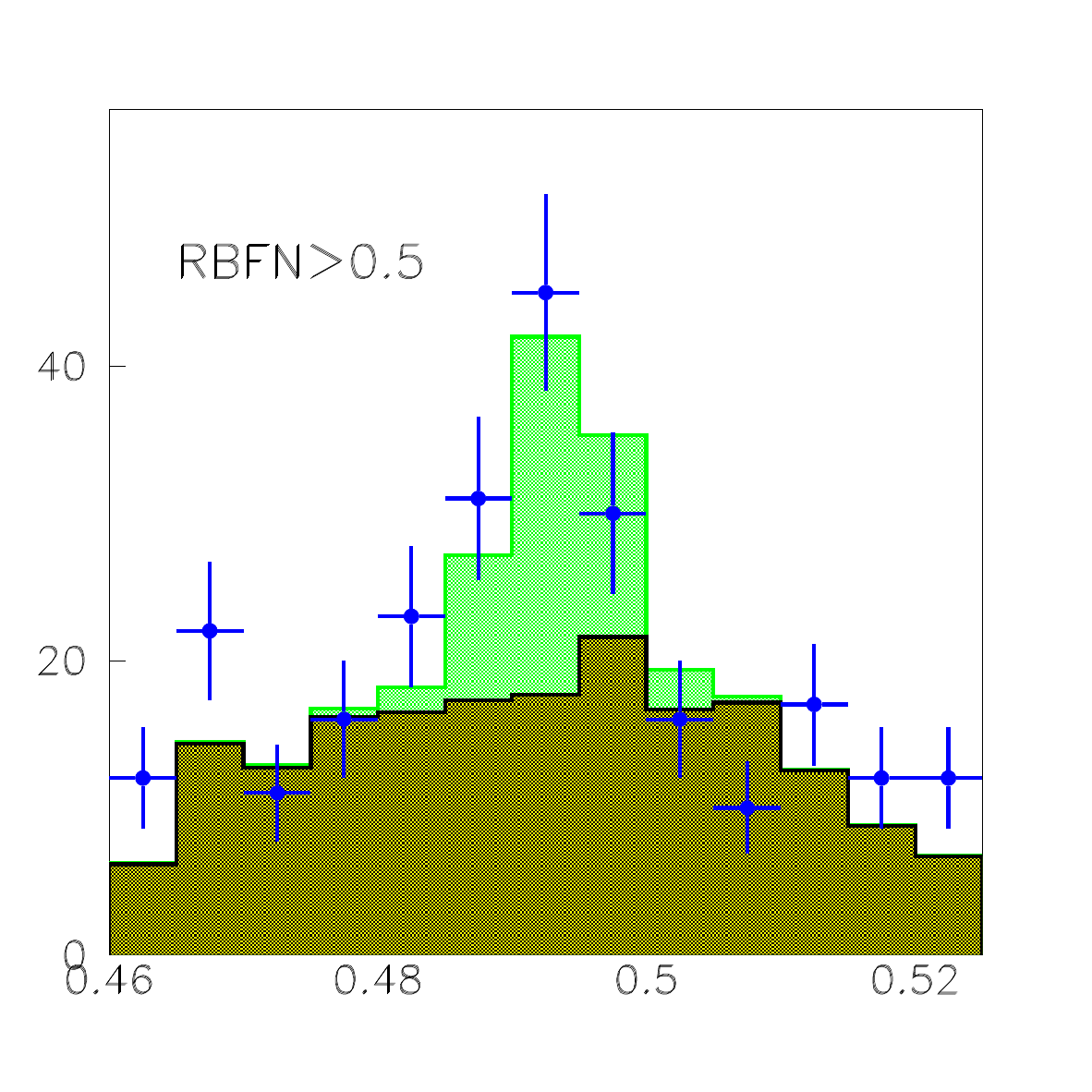}
\includegraphics[width=0.3\textwidth]{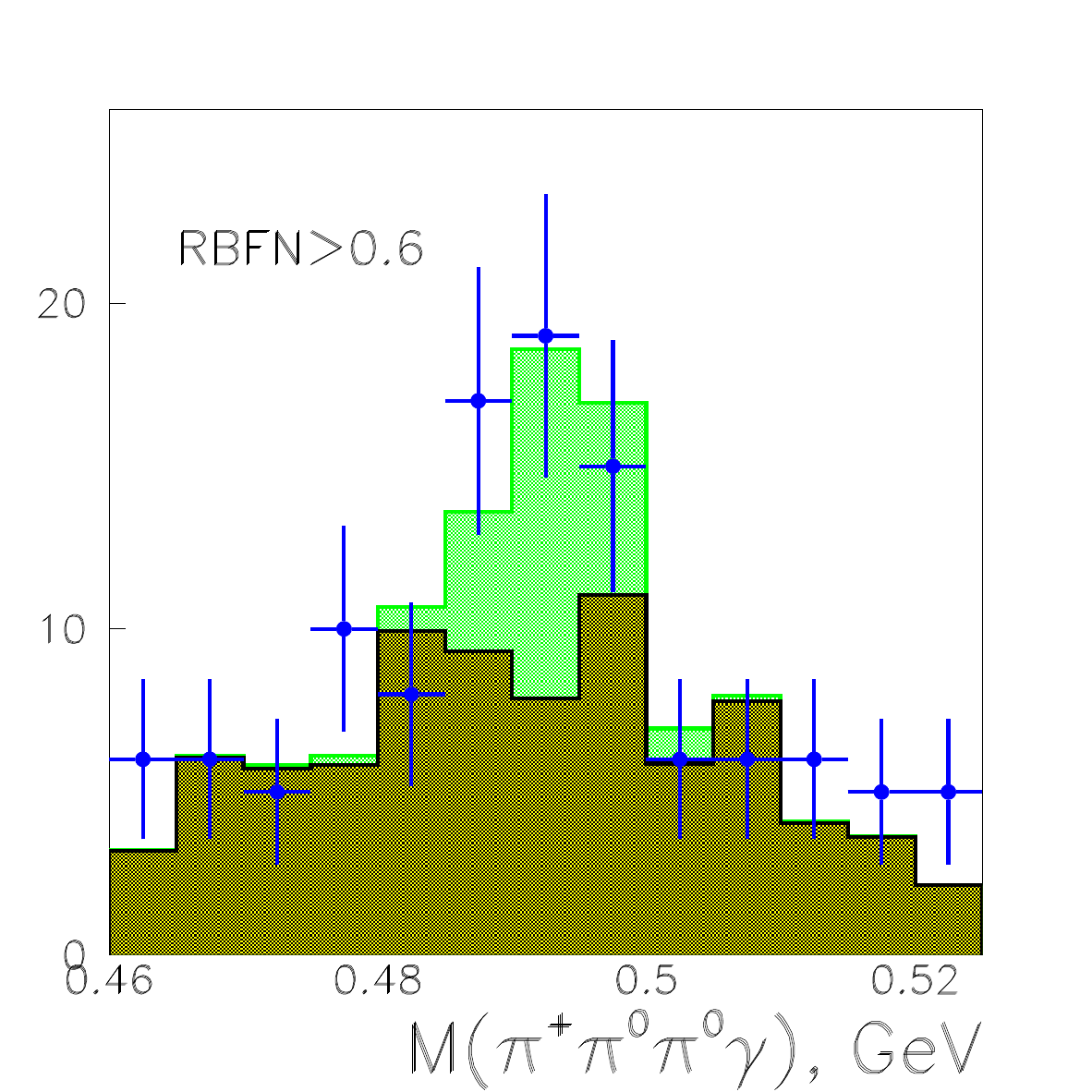}
\includegraphics[width=0.3\textwidth]{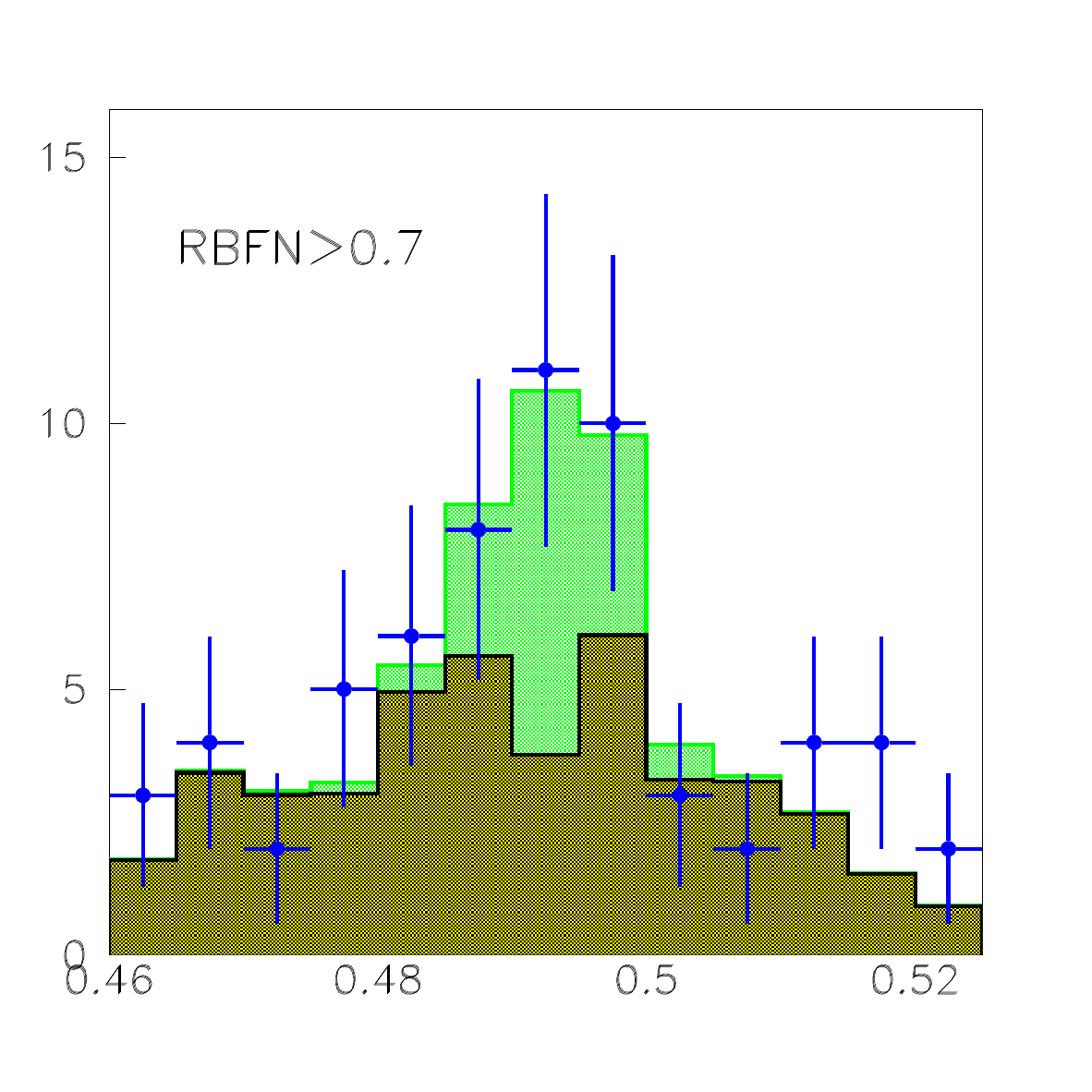}
\caption{\label{bkg-sub} Mass spectra for different cuts on RBFN output; data are the points with error bars,
  the MC background and signal are the dark and light histograms respectively.}
\end{center}
\end{figure*}

The mass spectra in fig.\ref{bkg-sub} are fitted in order to determine the number of events of the decay.
We tried three different background shape models to address the systematic errors possibly introduced by fit:

\begin{itemize}
\item MC shape: 
\begin{eqnarray}
  Data &=& \alpha \times MC(K^+ \rightarrow \pi^+\pi^0\pi^0\gamma) +    \nonumber \beta \times MC(K^+ \rightarrow \pi^+\pi^0\pi^0)   \nonumber 
\end{eqnarray}
with free scaling parameters $\alpha, \beta$;

\item Gaussian shape + second order polynomial with peak position and width fixed at their MC values;
\item Gaussian shape + second order polynomial, peak position and width being free parameters.

\end{itemize}

The numbers for diferent fits came out pretty close (Table \ref{fit-tabl-cum}).

\begin{table}[!h]
\caption{\label{fit-tabl-cum} Number of events in the peak for different RBFN thresholds and background models.}
\begin{center}
\centering
\begin{tabular}{ |c|c|c|c|} 
 \hline
   RBFN $>$ & MC & G+P2                 & G+P2   \\ 
              &    & $m,\sigma$ fixed     & $m,\sigma$ free   \\ 
 \hline
 0.5  & $52.8 \pm 11.7$ &  $63.2 \pm 11.7$ & $63 \pm 14.2$  \\ 
 \hline
 0.6  & $24.3 \pm 8$ &  $30.7 \pm 8$ & $30.7 \pm 9$ \\ 
 \hline
 0.7  & $14.5 \pm 6$   &  $18.9 \pm 5.9$ & $19.9 \pm 7.5$  \\ 
 \hline
\end{tabular}
\end{center}
\end{table}

\begin{table}[!h]
\caption{\label{fit-tabl-ind} Number of events in the peak for independent RBFN output ranges.}
\begin{center}
\begin{tabular}{ |c|c|c|c|} 
 \hline
  RBFN & MC & Gauss+P2                 & Gauss+P2   \\ 
                 &    & $m,\sigma$ fixed     & $m,\sigma$ free  \\ 
 \hline
 0.5-0.6  & $31.4 \pm 8.5$ &  $26.2 \pm 7.7$ & $28.1 \pm 10.3$  \\ 
 \hline
 0.6-0.7  & $8.1 \pm 5$ &  $9.7 \pm 4.5$ & $10.8 \pm 5.8$   \\ 
 \hline
 $>0.7$  & $14.5 \pm 6$   &  $18.9 \pm 5.9$ & $19.9 \pm 7.5$  \\ 
 \hline
\end{tabular}
\end{center}
\end{table}

\section{Branching ratio}

\begin{figure}
\begin{center}
\includegraphics[width=0.35\textwidth]{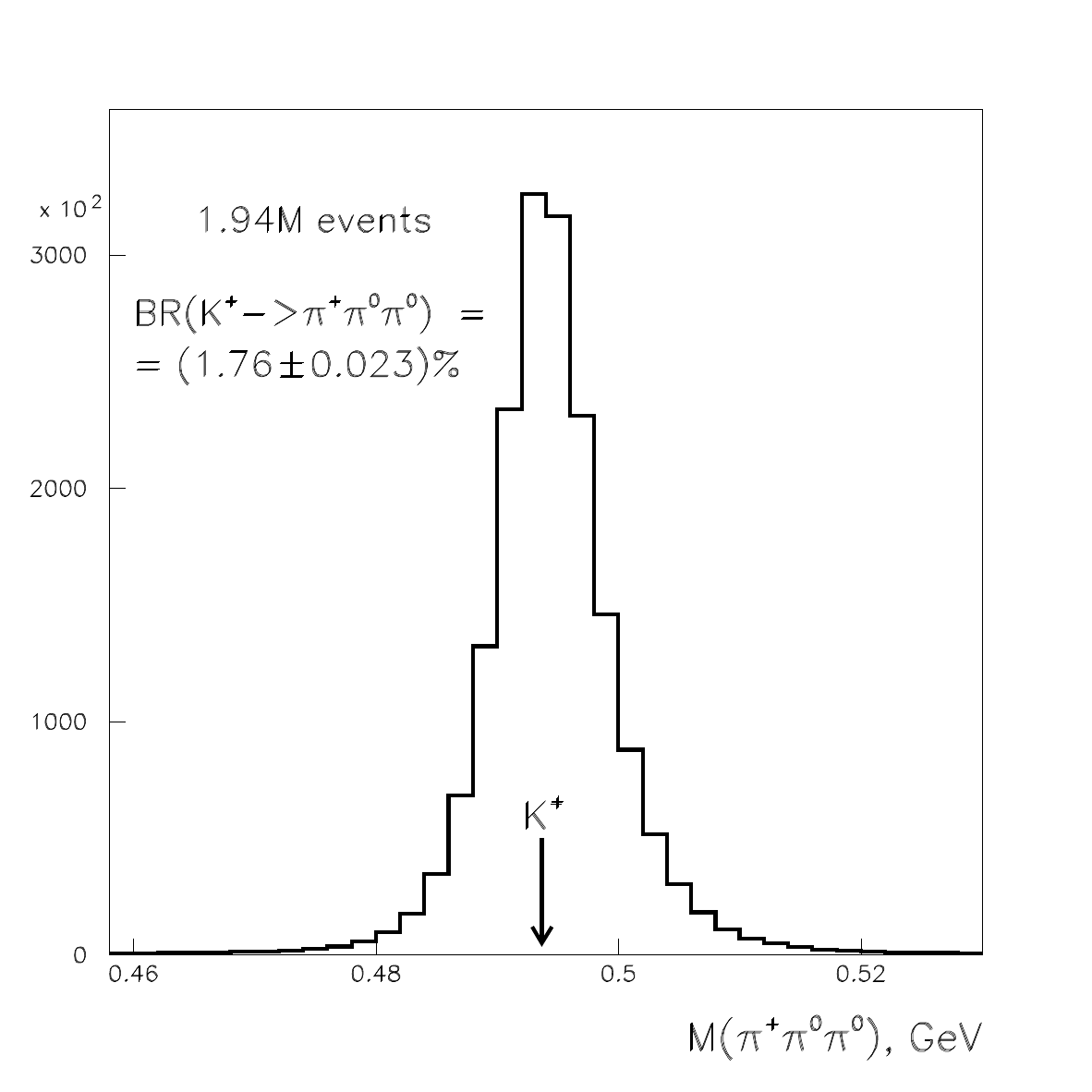}
\caption{\label{norm-peak} Observation of $K^{+}\to\pi^{+}\pi^{0}\pi^{0}$ decay used for normalization.}
\end{center}
\end{figure}

The decay of similar topology, $K^{+}\to\pi^{+}\pi^{0}\pi^{0}$, used for normalization to cancel out most of uncertainties in efficiency
calculation. We have about $2\times 10^6$ of those
at hand with no considerable background (fig.\ref{norm-peak}). The detection efficiencies for both decays obtained through MC simulation.

We obtain the mass spectra in 3 mutually independent intervals: $0.5<RBFN<0.6, 0.6<RBFN<0.7, RBFN>0.7$ by subtracting spectra in
fig.\ref{bkg-sub}. 
Each of these is fitted to 3 background models (Table \ref{fit-tabl-ind}). Finally we fit 9 numbers obtained this way
to calculate the BR:

\begin{equation}
\sum_{i=1}^9 \frac { \left( n_i - A\times BR \times \epsilon_i \right)^2} {\sigma_i^2} \to min
\label{BR-fit}
\end{equation}
where $n_i,\sigma_i,\epsilon_i$ stand for numbers of events in the peak, their errors and detection efficiencies respectively,
the sum runs over RBFN intervals and background models, and BR is the free fit parameter.
The normalization constant A is given by
\begin{equation}
A = \frac{\epsilon_{\pi^{+}\pi^{0}\pi^{0}}}{n_{\pi^{+}\pi^{0}\pi^{0}}} \times BR_{\pi^{+}\pi^{0}\pi^{0}}
\label{A}
\end{equation}
where $\epsilon_{\pi^{+}\pi^{0}\pi^{0}},n_{\pi^{+}\pi^{0}\pi^{0}}$ and $BR_{\pi^{+}\pi^{0}\pi^{0}}$
are detection efficiency, observed number of events and branching ratio of the normalization decay respectively.
The result of this fit is
\begin{equation}
BR(K^{+}\to\pi^{+}\pi^{0}\pi^{0}\gamma) = (3.78 \pm 0.8) \times 10^{-6}, \quad E_{\gamma}^*>10MeV
\label{BR}
\end{equation}
$\chi^2/N_{DF}=3.5/8$ shows no evidence of systematic error due to RBFN threshold or background model.

\section{Systematic errors}

All known systematic errors are well below $20\%$ statistical error.

\begin{itemize}
\item Mass spectra fitting. Instead of fitting the spectra in indepent RBFN intervals 
  we fit cumulative spectra in fig.\ref{bkg-sub} and then obtain the numbers $n_i$ in (\ref{BR-fit}) by subtraction the rows
  of Table \ref{fit-tabl-cum}.
  These 9 numbers come out different from those listed in
   table\ref{fit-tabl-ind}. Then derive the BR per (\ref{BR-fit}). The result obtained this way is $BR=(3.39 \pm 0.76) \times 10^{-6}$,
   $11.5\%$ less than (\ref{BR}). So we assign the systematic error due to this source to $\sigma_{syst(fit)} = 11.5\%/\sqrt{2}=8\%$.

   
\item Overall normalization uncertainty evaluated to $4\%$ including selection criteria and $BR(K^{+}\to\pi^{+}\pi^{0}\pi^{0})$.
  \item $\gamma$ detection threshold varied from 0.4 to 0.6 GeV, $4\%$ variation in BR.
  \item GAMS threshold curve. The triggering probability raises gradually from 0 to 1 with the GAMS energy deposition increase.
    The MC study showed detection efficiencies for both decays being insensitive to particular shape of this curve down to $1\%$.
\end{itemize}
All these sources, being added in quadrature, result in $\sigma_{syst} = 10\% \approx 0.4 \times 10^{-6}$. So the overall error is
driven by limited statistics.


\section{$\gamma$ spectrum}

\begin{figure}
\begin{center}
\includegraphics[width=0.35\textwidth]{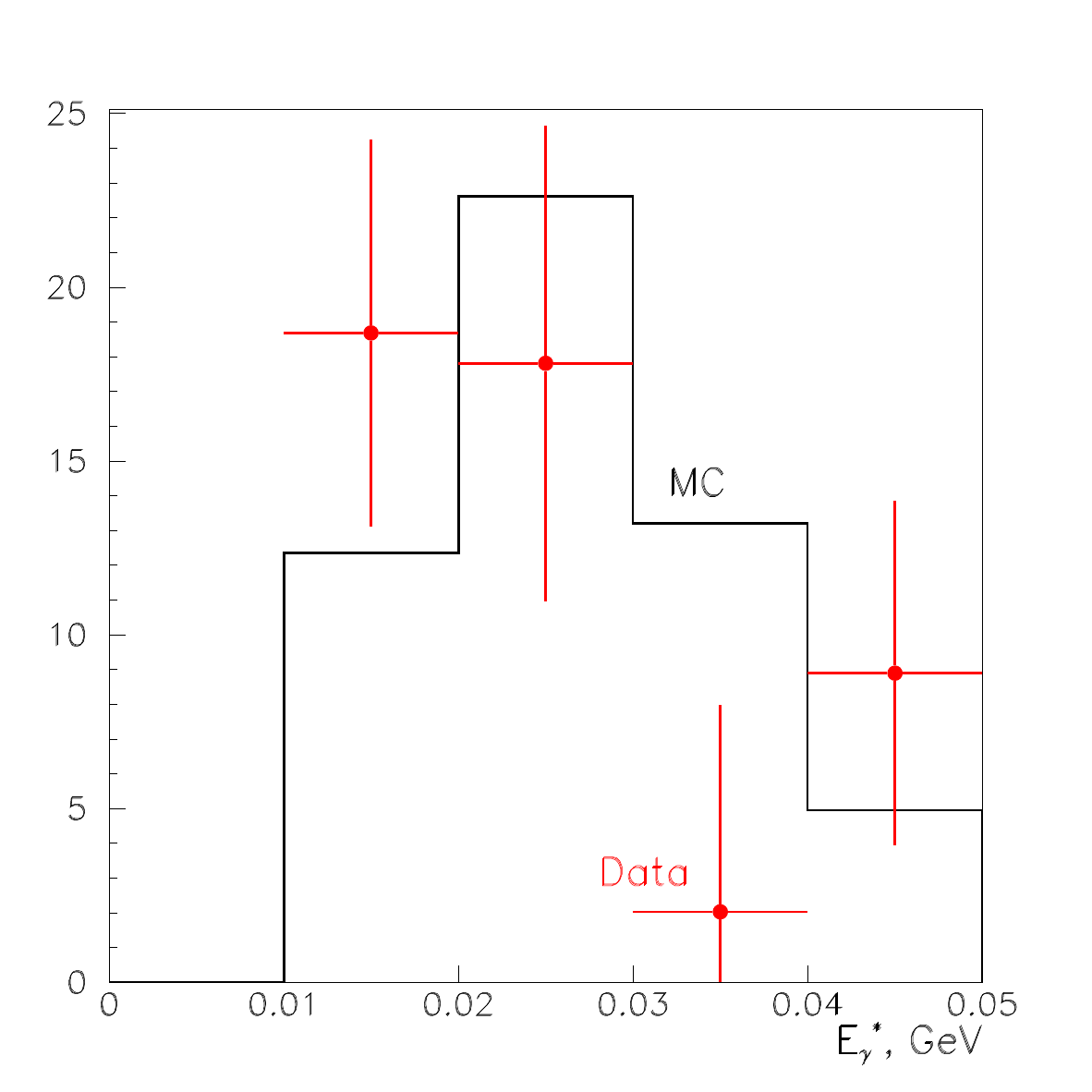}
\caption{\label{e5cm} Energy spectrum of $\gamma$ in $K^+$ rest frame.}
\end{center}
\end{figure}

The area around $K^+$ mass $(0.485<m(\pi^+\pi^0\pi^0\gamma)<0.505)$ of the spectrum in the leftmost frame of fig.\ref{bkg-sub} (RBFN>0.5)
selected for this study. Scaled MC $\gamma$ spectrum was then subtracted from the $\gamma$ spectrum of selected events:

\begin{equation}
\frac{dn}{dE_{\gamma}^*} = \left( \frac{dn}{dE_{\gamma}^*} \right)_{exp} - \beta \times \left( \frac{dn}{dE_{\gamma}^*} \right)_{bkg MC},
\label{spektr}
\end{equation}
with scaling factor $\beta$ obtained from mass spectrum fit. $E_{\gamma}^*$ is the $\gamma$ energy in $K^+$ rest frame.
The resulting spectrum agrees with ChPT prediction although the errors are large (fig.\ref{e5cm}).

\section*{Conclusions}
The OKA data are analyzed in search for $K^{+}\to\pi^{+}\pi^{0}\pi^{0}\gamma$ decay.
The major background source, the decay \\
$K^{+}$$\to$$\pi^{+}\pi^{0}\pi^{0}$,
is $\approx 5000$ times more intense than the radiative decay in question.
The RBFN neural network is employed to suppress the background down to Signal:Noise$\approx 1:1$ level; about 60 events of the decay
observed. The branching ratio obtained by normalization to the decay of similar topology $K^{+}\to\pi^{+}\pi^{0}\pi^0$,
$BR(K^{+}\to\pi^{+}\pi^{0}\pi^0\gamma) = \\ (4.1 \pm  0.9(stat) \pm  0.4(syst)) \times 10^{-6} (E_{\gamma}^*>10MeV)$
agrees with ChPT prediction of $3.76 \times 10^{-6}$. 
The $\gamma$ energy spectrum is also in agreement with ChPT although the errors are large.
The observation of the decay proves feasibility of its detailed study on a larger statistical sample.

%

\subsection*{Acknowledgements}

We express our gratitude to our colleagues in the accelerator department for the good performance of the U-70 during data taking; 
to colleagues from the beam department for the stable operation of the 21K beam line, including RF-deflectors, and to colleagues 
from the engineering physics department for the operation of the cryogenic system of the RF-deflectors.\\
This work was supported by the RSCF grant {\it{N\textsuperscript{\underline{\scriptsize o}}}22-12-0051}.

\end{document}